\documentclass[12pt]{article}
\usepackage{epsfig}
\usepackage{amssymb}
\usepackage{overpic}
\usepackage{multirow}
\usepackage{dcolumn}

\def\beq{\begin{equation}}
\def\eeq#1{\label{#1}\end{equation}}
\def\eeqn{\end{equation}}

\def\beqa{\begin{eqnarray}}
\def\eeqa#1{\label{#1}\end{eqnarray}}
\def\eeqan{\end{eqnarray}}

\let\bar=\overbar

\def\etal{{\it et al.}}

\def\Dslash{\not{\hbox{\kern-4pt $D$}}}
\def\dslash{\not{\hbox{\kern-2pt $\del$}}}

\def\msb{{\bar{\ssstyle M \kern -1pt S}}}

\def\Title#1{\begin{center} {\Large {\bf #1} } \end{center}}

\newcommand{\ttbar}{\ensuremath{t\bar{t}}}

\newcommand{\etmissx}{\ensuremath{E \kern-0.6em\slash_{\rm x}}}
\newcommand{\etmissy}{\ensuremath{E \kern-0.6em\slash_{\rm y}}}

\newcommand{\ljets}{\ensuremath{\ell+{\rm jets}}}
\newcommand{\gtop}{\ensuremath{\Gamma_t}}
\newcommand{\gtwb}{\ensuremath{\Gamma_{t\rightarrow Wb}}}
\newcommand{\btwb}{\ensuremath{\mathcal B_{t\rightarrow Wb}}}
\newcommand{\sigt}{\ensuremath{\sigma_{t-{\rm channel}}}}
\newcommand{\afb}{\ensuremath{A_{\rm fb}}}

\newcommand{\del}{\partial}

\newcommand{\GeV}{\ensuremath{\textnormal{GeV}}}
\newcommand{\TeV}{\ensuremath{\textnormal{TeV}}}

\newcommand{\fb}{\ensuremath{{\rm fb}^{-1}}}

\newcommand{\mttbar}{\ensuremath{m_{t\bar t}}}
\newcommand{\absdy}{\ensuremath{|\Delta y|}}

\begin{document}

\Title{Recent top physics results from the D0 experiment}

\bigskip\bigskip

%+\addtocontents{toc}{{\it D. Reggiano}}
%+\label{ReggianoStart}

\begin{raggedright}  
{\it Oleg Brandt\index{Brandt, O.} on behalf of the D0 collaboration\\
II. Physikalisches Institut\\
Georg August Universit\"at G\"ottingen\\
D-37077 G\"ottingen, Germany}
\bigskip
\end{raggedright}

\begin{abstract}
We review recent measurements of the properties of the top quark by the D0 experiment: the decay width of the top quark, the CKM matrix element~$V_{tb}$, the helicity of the $W$ boson, anomalous couplings at the $Wtb$ vertex, violation of invariance under Lorentz transformations, and the asymmetry of $\ttbar$ production due to the strong colour charge. The measurements are performed using data samples of up to 5.4~\fb\ acquired by the D0 experiment in Run II of the Fermilab Tevatron $p\bar p$ collider at a centre-of-mass energy of $\sqrt s=1.96~\TeV$.
\vspace{2mm}\\
PACS {\tt 14.65.Ha} -- Top quarks.
\end{abstract}

%\section{Introduction}
The pair-production of the top quark was discovered in 1995 by the CDF and D0 experiments~\cite{bib:topdiscovery} at the Fermilab Tevatron proton-antiproton collider. Observation of the electroweak (EW) production of single top quarks was presented only 3 years ago~\cite{bib:singletopdiscovery}. The large top quark mass of $173.2\pm0.9$~\GeV~\cite{bib:mtop} and the resulting Yukawa coupling of about $0.996\pm0.006$ suggest that the top quark could play a crucial role in EW symmetry breaking. Precise measurements of the properties of the top quark provide a crucial test of the consistency of the standard model (SM) and could hint at physics beyond the SM. The full listing of top quark measurements by the D0 experiment can be found in~Ref.~\cite{bib:topresd0}.

At the Tevatron, top quarks are mostly produced in pairs via the strong interaction. %, in about 85\% of the cases via $q\bar q'$ annihilation and in about 15\% via gluon-gluon fusion. 
In the framework of the SM, the top quark decays to a $W$ boson and a $b$ quark nearly 100\% of the time, resulting in a $W^+W^-b\bar b$ final state from top quark pair production.
Thus, $\ttbar$ events are classified according to the $W$ boson decay channels as ``dileptonic'', ``all--jets'', or ``lepton+jets''. The EW production of single top quarks is classified via $s$ and $t$ channel, as well as associated $Wt$ production.

\bigskip
%\section{Measurement of the decay width of the top quark}
We extract the total decay width of the top quark~\cite{bib:width} from the partial decay width $\gtwb$, measured using the $t$-channel cross section for single top quark production~\cite{bib:singletop}, and from the branching fraction $\btwb$, measured in $\ttbar$ events~\cite{bib:ttbar}, from up to 5.4~\fb\ of data. This extraction is made under the assumption that the EW coupling in top quark production is identical to that in the decay, and using the next-to-leading order~(NLO) calculation of the {\it ratio} $\gtwb^{\rm SM}/\sigt^{\rm SM}$, i.e.~$\gtop=\frac{\sigt}{\btwb}\times\frac{\gtwb^{\rm SM}}{\sigt^{\rm SM}}$.
Properly taking into account all systematic uncertainties and their correlations among the measurements of \gtwb\ and \sigt, we find $\gtop=2.00^{+0.47}_{-0.43}~\GeV$, which translates into a top-quark lifetime of $\tau_t=(3.3^{+0.9}_{-0.6})\times10^{-25}$~s, in agreement with the SM expectation. This constitutes the world's most precise indirect determination of \gtop\ to date.
Furthermore, we use the $t$-channel discriminant of the above measurement to extract a limit 
%on the CKM matrix element $V_{tb}$ 
of $|V_{tb}|>0.81$ at 95\%~C.L., without the commonly made assumptions that $t\to Wb$ exclusively, or on the relative $s$ and $t$ channel rates.

\bigskip
%\section{Measurement of $W$ boson helicity in $\mathbf{t\bar t}$ events}
In the SM, the top quark decays into a $W$ boson and a $b$ quark with a probability of $>99.8\%$, where the on-shell $W$ boson can be in a left-handed, longitudinal, and right-handed helicity state. A NLO calculation in the SM predicts $f_-=0.301,\,f_0=0.698,$ and $f_+=4.1\times10^{-4}$, respectively. A deviation from the SM expectation could indicate a contribution from new physics. We simultaneously measured the $f_0$ and $f_+$ helicity fractions in the dilepton and \ljets\ final states using 5.4~\fb\ of data~\cite{bib:whel_d0}. A model-independent fit to the distribution in $\cos\theta^*$, where $\theta^*$ is the angle between the three-momentum of the top quark and the down-type fermion in the $W$ boson rest frame, yields $f_0=0.67\pm0.10$ and $f_+=0.02\pm0.05$, in agreement with the SM expectation. The combination with measurements by CDF in the dilepton and \ljets\ final states using up to 4.8~\fb\ yields $f_0=0.72\pm0.08$ and $f_+=-0.03\pm0.46$~\cite{bib:whel}.

\bigskip
% anomalous couplings at the Wtb vertex
The SM provides a purely left-handed vector coupling at the $Wtb$ vertex, while the most general and lowest-dimension effective Lagrangian 
%\[
%\mathcal{L}=\frac g{\sqrt 2}\bar b\gamma^\mu V_{tb}(f^L_V P_L + f^R_V P_R)tW_\mu^-
%- \frac g{\sqrt 2}\bar b\frac{i\sigma^{\mu\nu}q_\nu V_{tb}}{M_W} (f^L_T P_L + f^R_T P_R)tW_\mu^-
%+ h.c.\,,
%\]
allows a right-handed vector coupling $f^R_V$ as well as tensor couplings $f^L_T$ and $f^R_T$. We extracted limits on those anomalous couplings from single top production in 5.4~\fb\ using both shapes of kinematic distributions as well as the overall and $s$ versus $t$~channel event rates~\cite{bib:anomal_st}. This was done under the assumption of real, i.e.\ $CP$-conserving couplings and a spin~1/2\linebreak top quark predominantly decaying to $Wb$. 
%A~Bayesian NN in three jet bins and two $b$ tag bins was trained against three models in which one non-vanishing anomalous coupling was allowed at a time. 
The results are shown in Table~\ref{tab:anomal}.\\
Furthermore, we exploit that anomalous couplings at the $Wtb$ vertex will alter the rates of the three helicity states of $W$ bosons in $\ttbar$ decays, and combine the above analysis with the one in Ref.~\cite{bib:whel_d0}
%, removing overlap between the two selections and properly accounting for correlations, 
to obtain improved limits~\cite{bib:anomal}, shown in Table~\ref{tab:anomal}.

\begin{table}[htbp]
\centering
\begin{tabular}{l|ccc}
\hline
Scenario & $W$ helicity only & single top only & combination \\
\hline 
$|f_V^R|^2$ & 0.62     & 0.89     & 0.30 \\
$|f_T^L|^2$ & 0.14     & 0.07     & 0.05 \\
$|f_T^R|^2$ & 0.18     & 0.18     & 0.12 \\
\hline
\end{tabular}
\caption{
  \label{tab:anomal}
  Observed upper limits on anomalous $Wtb$ couplings at 95\% C.L. from $W$~boson helicity assuming $f_V^L=1$, from the analysis of single top events, and their combination, for which no assumption on $f_V^L$ is made, using 5.4~\fb.
}
\end{table}

\bigskip
%LIV analysis
Invariance under Lorentz transformations is a fundamental property of the SM. 
%It was already thoroughly tested in the leptonic sector, for the first two quark generations, and in the $b$ quark system. 
We performed a search for Lorentz invariance violation (LIV) by examining the \ttbar\ production cross section in \ljets\ final states using 5.3~\fb~\cite{bib:lorentz}. We quantified LIV in the top sector using in the SM Extension (SME) formalism~\cite{bib:sme}, which parametrises the amount of LIV in terms of bilinear coefficients and
%. Non-vanishing coefficients phenomenologically 
results in a \ttbar\ production rate which is modulated in full or half-units of a siderial day due to the rotation of the Earth. Thus, we
%We use the time stamp of the \ttbar\ candidate events and 
investigate the \ttbar\ production rate in 12 bins of the siderial phase, normalised by the recorded luminosity per bin. Our results are consistent with the SM hypothesis, and we proceed to set limits on LIV~\cite{bib:lorentz}.
%, which cannot be listed here for space reasons.

\bigskip
%\section{Measurement of the strong colour charge production asymmetry in $\mathbf{t\bar t}$ events}
In the SM, the pair production of top quarks in $p\bar p$ collisions, a $CP$ eigenstate, is symmetric at LO under charge conjugation. NLO calculations %within the SM 
predict a small forward-backward asymmetry $\afb$ of the order of 5\% in the $\ttbar$ rest frame. It is due to a negative contribution from the interference of diagrams for initial and final state radiation, and a (larger) positive contribution from the interference of box and tree-level diagrams. This experimental situation is unique to the Tevatron. 
%because the initial $p\bar p$ state is a $CP$ eigenstate, and because $\ttbar$ production is almost at threshold. 
A convenient observable for the Tevatron is $\afb\equiv\frac{N^{\Delta y>0}-N^{\Delta y<0}}{N^{\Delta y>0}+N^{\Delta y<0}},$ where $\Delta y\equiv y_t-y_{\bar t}$, $y_t$~($y_{\bar t}$) is the rapidity of the $t$~($\bar t$) quark. 
%, and $y=\frac12\ln\frac{E+p_z}{E-p_z}$. 
Another common observable does not depend on a full reconstruction of the \ttbar\ system: $\afb^\ell\equiv\frac{N^{q_\ell y_\ell>0}-N^{q_\ell y_\ell<0}}{N^{q_\ell y_\ell>0}+N^{q_\ell y_\ell<0}}$, where $q_\ell$ is the lepton charge.
%, which offers superior experimental resolution as it but is statistically less sensitive than the $\afb$ observable.

\begin{figure}[htb]
\centering
\begin{overpic}[width=0.48\textwidth]{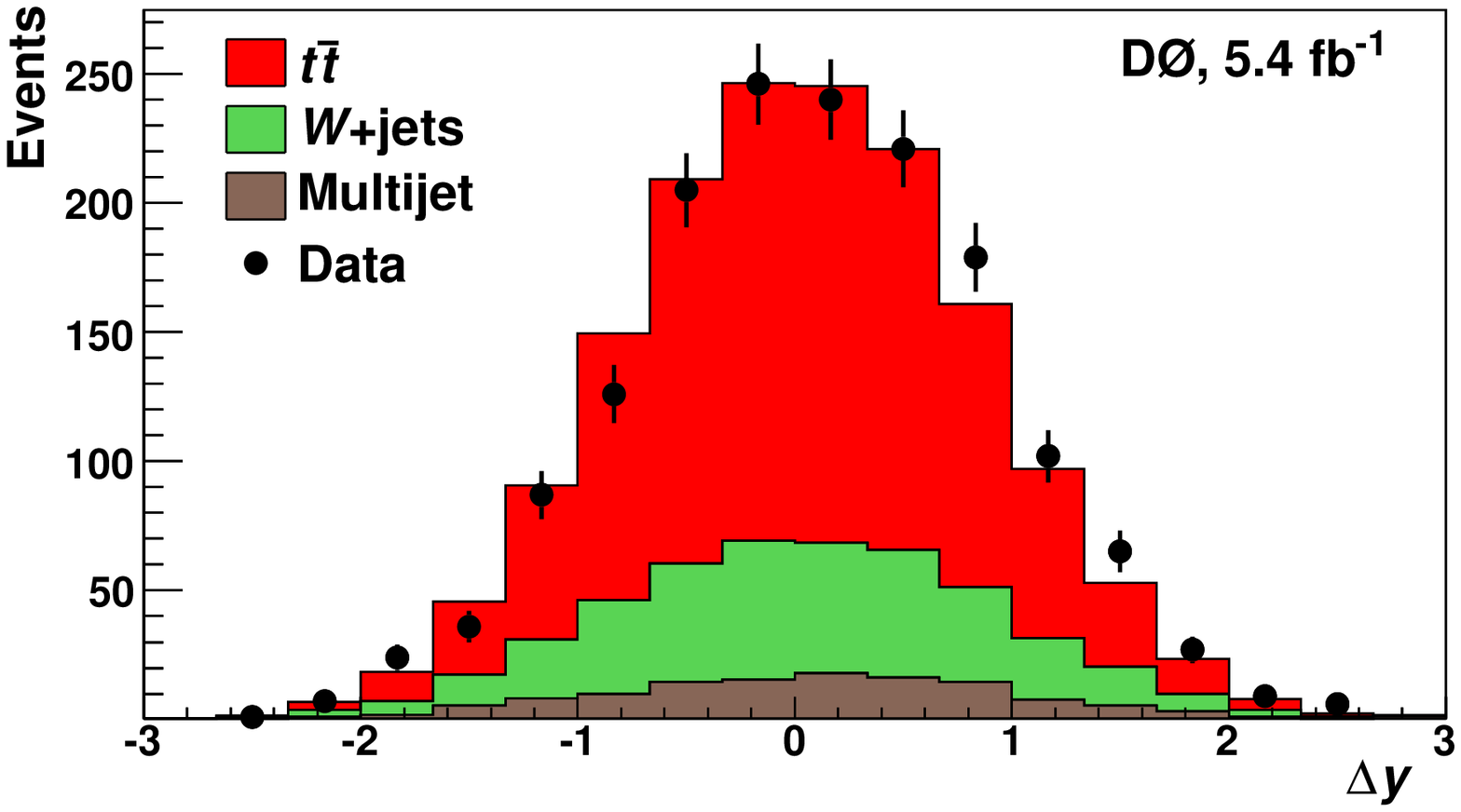}
\put(-1,2){(a)}
\end{overpic}
\quad
\begin{overpic}[width=0.48\textwidth]{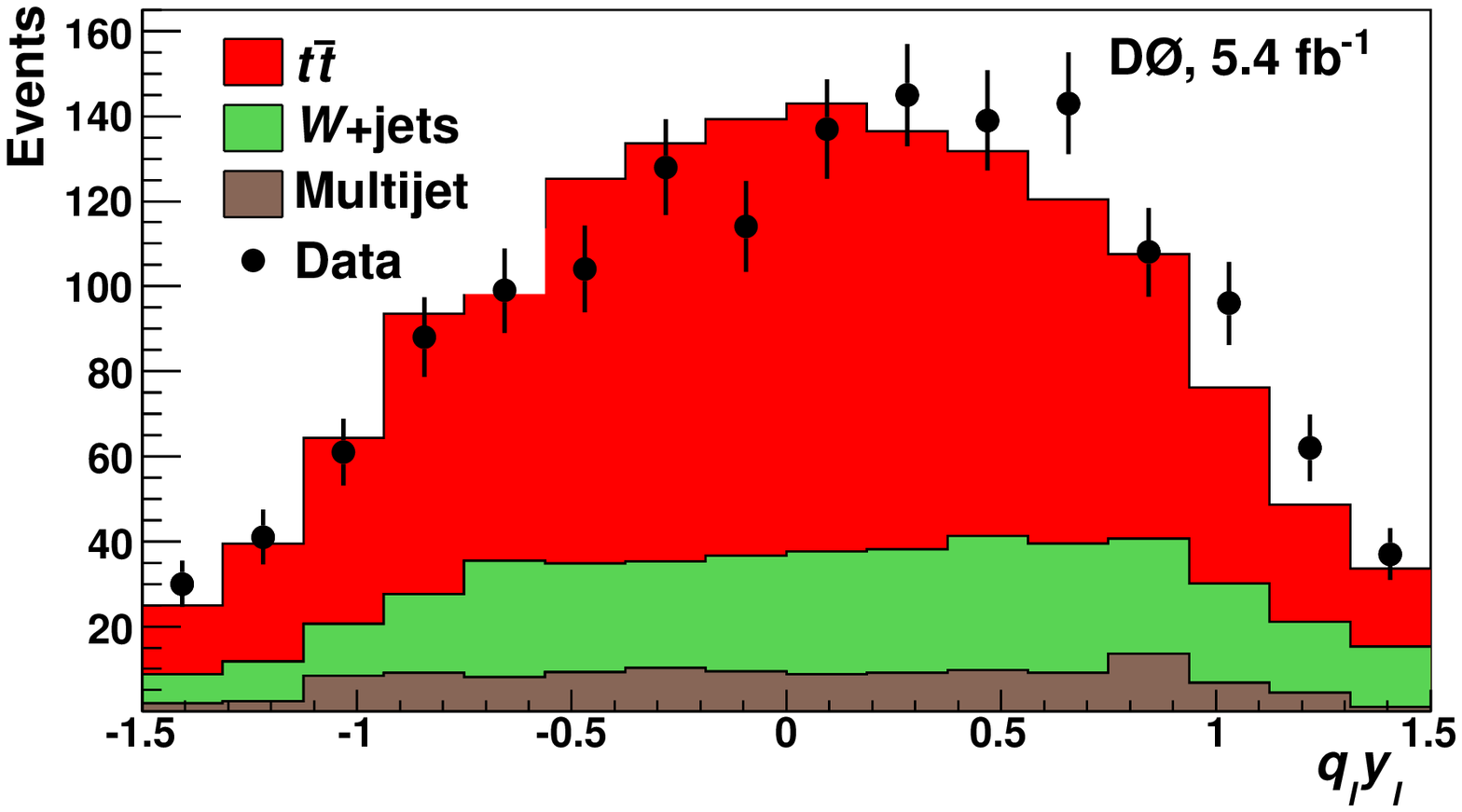}
\put(-1,2){(b)}
\end{overpic}
\caption{
\label{fig:afb}
{\bf(a)} The reconstructed distribution in $\Delta y$ in \ljets\ final states using 5.4~\fb. 
%Bin width correspond to about half of the detector resolution in $\Delta y$.
{\bf(b)} The reconstructed distribution in charge-signed lepton rapidity $q_\ell y_\ell$.
}
\end{figure}

%\begin{table}[htbp]
%\centering
%\begin{tabular}{l*{2}{D{,}{\,\pm\,}{-1}}}
%\hline
%  &  \multicolumn{2}{c}{\afb\ (\%)}\\
%Subsample  & \multicolumn{1}{c}{\hspace{1ex}Data} & \multicolumn{1}{c}{\hspace{1em}{\sc mc@nlo}} \\ 
%\hline 
%$\mttbar<450\GeV$ &  7.8 , 4.8 & 1.3 , 0.6  \\   
%$\mttbar>450\GeV$ & 11.5 , 6.0 & 4.3 , 1.3  \\
%$\absdy<1.0$      &  6.1 , 4.1 & 1.4 , 0.6  \\
%$\absdy>1.0$      & 21.3 , 9.7 & 6.3 , 1.6 \\
%\hline
%\end{tabular}
%\caption{
%  \label{tab:afb}
%  Reconstructed \afb\ by subsample  in \ljets\ final states using 5.4~\fb.
%}
%\end{table}

%
We measured \afb\ in the $\ttbar$ rest frame in \ljets\ final states on a dataset corresponding to 5.4~\fb\ using $\ttbar$ event candidates fully reconstructed with a kinematic fitter, and found $\afb=9.2\%\pm3.7\%$ at the reconstruction level~\cite{bib:afb}. Our result, shown in Fig.~\ref{fig:afb}~(a), is about 1.9 standard deviations (SD) away from the {\sc mc@nlo}~\cite{bib:mcnlo} prediction of $2.4\pm0.7\%$. After correcting for detector acceptance and resolution we find $\afb=19.6\pm6.5\%$, 2.4~SD away from the {\sc mc@nlo} prediction of $5.0\pm0.1\%$. In addtion, we measured \afb\ in various subsamples
defined by $\mttbar\lessgtr450~\GeV$ and by $\absdy\lessgtr1.0$.
%which are as shown in Table~\ref{tab:afb}. 
We do not find any statistically significant dependencies. Furthermore, we have measured the lepton-based asymmetry and find $\afb^{\ell}=14.2\pm3.8\%$ and $15.2\pm4.0\%$ at reconstruction and parton level, respectively, while {\sc mc@nlo} predicts  $\afb^{\ell}=0.8\pm0.6\%$ and $2.1\pm0.1\%$. Our results display some tension with the NLO SM prediction. This may indicate a contribution from new physics, but may as well be due to contributions at higher orders in $\alpha_s$ within the SM. 
%Several mechanisms originating from new physics contributions have been suggested to explain this discrepancy, however, it has been pointed out that non-vanishing and acceptance-dependent contributions at higher orders in $\alpha_s$ within the SM could play an important role in understanding these findings.

\bigskip
%\section{Conclusions}
I presented recent measurements of key properties of the top quark by the D0 experiment, most in good agreement with SM expectations. The forward-backward  asymmetry $\afb$ of $\ttbar$ production displays tension between the measurement and the SM NLO calculations. We look forward to updates with the full dataset of 9.7~\fb\ in the near future.
I would like to thank my fellow D0 collaborators and also 
%for their help in preparing this article. I also thank 
the staffs at Fermilab and collaborating institutions, as well as the D0 funding agencies.

\end{document}